\documentclass[showpacs,prl,twocolumn]{revtex4}
\usepackage{graphicx}
\begin{document}
\title{Theory of pattern-formation  of
metallic microparticles in poorly conducting liquid}
\author{I. S. Aranson$^1$ and M. V. Sapozhnikov$^{1,2}$}
\affiliation{$^{1}$Materials Science Division, Argonne National
Laboratory, 9700 South Cass Avenue, Argonne, IL 60439 \\
$^{2}$Institute for Physics of Microstructures, Russian Academy of Sciences, GSP-105,
Nizhny Novgorod, 603000, Russia }
\date{\today}
\begin{abstract}
We develop  continuum theory of self-assembly  and pattern
formation  in metallic microparticles immersed in a poorly
conducting liquid in DC electric field. The theory is formulated
in terms of two conservation laws for the densities of immobile
particles (precipitate) and bouncing particles (gas) coupled to
the Navier-Stokes equation for the liquid. This theory
successfully reproduces correct  topology of the phase diagram and
primary patterns observed in the experiment [Sapozhnikov et al,
Phys. Rev. Lett. {\bf 90}, 114301 (2003)]: static crystals and
honeycombs and dynamic pulsating rings and rotating  multi-petal
vortices.
\end{abstract}
 \maketitle

Electromagnetic manipulation  and assembly of small particles in
electrolytes has constitutes one of the great hopes for
nanotechnology \cite{hayward} and new generation of microfluidic
devices \cite{chang}. Many industrial technologies face the
challenge of handling such single- or multi-component micro and
nano- size ensembles. The dynamics of conducting microparticles in
electric field in the air was studied in \cite{ar1,ar2}. Phase
transitions and clustering instability of the electrostatically
driven conducting microparticles  were found. In our recent work
we reported new dynamic phenomena occurring with microparticles in
poorly conducting liquid subject to strong electric field (up to
20 kV/cm) \cite{sap}. It was shown that small metallic particles
immersed in a toluene-ethanol mixture in DC electric field form a
rich variety of novel phases. These phases include static
honeycombs and two-dimensional crystals; dynamic multi-petal
vortices and pulsating rings. The phenomena were attributed to
interaction between particles and electro-hydrodynamic (EHD)
flows.

In this Letter we develop  theory of pattern formation and
self-assembly of metallic microparticles in a poorly conducting
liquid placed between two horizontal planar electrodes. The theory
is formulated in terms of two depth-averaged conservation laws for
the densities of immobile particles (precipitate) and bouncing
particles (gas) coupled to the Navier-Stokes equation for vertical
component of velocity of the liquid $v_z$, whereas horizontal
velocity $v_\perp$ is obtained from the continuity condition. This
theory successfully reproduces topology of the phase diagram and
primary patterns observed in the experiment: static crystals and
honeycombs and dynamic pulsating rings and vortices \cite{sap}. In
the framework of our theory we demonstrate that the rotation of
clusters is the result of a symmetry-breaking instability leading
to formation of the travelling wave at the cluster perimeter.
% Our
%theory constitute substantial generalization and development of
%our previous work on coarsening in air-filled cell \cite{ar2}. The
%major difference with \cite{ar2} is explicit incorporation of the
%particulate transport by EHD flows.
%The model has many parameters which is
%mostly complexity of original physical system.

\begin{figure}[ptb]
\includegraphics[angle=-90,width=3.3in]{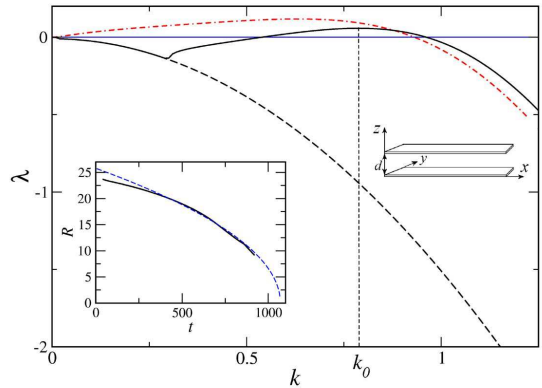}
\caption{Real part of $\lambda$ vs $k$ for three  regimes:
honeycomb (solid line, $E=-45, \beta=2, \kappa=-0.1 $),
coalescence (point-dashed line $E=-55, \beta=0.5, \kappa=0.1 $),
stable 2D crystal (dashed line, $E=50, \beta=2,\kappa=-0.1$).
Other parameters: $\zeta=0.02, \bar \rho=0.3, D_p=1,
\alpha_p=\alpha_g=-0.3, \nu=2, d_0=1, \mu_0=0$. Inset: Schematics
of experimental setup (right) and distance $R$ between  t-vortices
vs $t$ (solid line) for parameters of Figure \protect
\ref{figure3} (left). Dashed line shows fit $R \sim \sqrt{t_0
-t}$.} \label{figure1}
\end{figure}

{\it Model}. Schematics of experimental setting  is shown in
inset of  Figure \ref{figure1}. Our  studies show that two major
control parameters are the potential difference $\Delta U$ which
determines average electric field $E= -\Delta U/d$, $d$ is the
spacing between electrodes, and the concentration $c$ of the
additive (e.g. ethanol)  which  characterizes the conductivity of
the liquid. Following the analysis of Ref. \cite{ar2}, we describe
the evolution of particulate by the number density of precipitate
$\rho_p({\bf r},t)$ and bouncing particles (or gas) $\rho_g({\bf
r},t)$, where ${\bf r}=(x,y)$ are horizontal coordinates. All the
quantities are averaged over the vertical coordinate $z$. Since
the total number of particles $N = \int (\rho_p+\rho_g) dx dy$ is
conserved, the evolution of the particulate is described by the
conservation laws
\begin{eqnarray}
\partial_t \rho_p = \nabla {\bf J}_p + f \;,\;
\partial_t \rho_g = \nabla {\bf J}_g - f
\label{con_laws}
\end{eqnarray}
Here $J_{p,g}$ are the mass fluxes of precipitate and gas
respectively and the function $f$ describes gas/precipitate
conversion which depends on $\rho_{p,g}$, electric field $E$ and
local concentration $c$. The fluxes can be written as:
\begin{equation}
{\bf J}_{p,g}=D_{p,g} \nabla \rho_{p,g} + \alpha_{p,g}(E)  {\bf
v}_\perp  \rho_{p,g} (1- \beta(E)  \rho_{p,g}) \label{flux}
\end{equation}
where $D_{p,g}$ are precipitate/gas diffusivities.  The last term,
describing  particles advection  by fluid, is reminiscent of
Richardson-Zaki relation for drag force  frequently used in the
engineering literature \cite{richardson}. The factor $(1- \beta(E)
\rho_{p,g})$ describes saturation of the flux at large particle
densities $\rho \sim 1/\beta$. Experiments show that at the onset
of motion the maximum density  in the patterns such as honeycombs
is below sub-monolayer coverage whereas for large values of $E$
the particles form multi-layered structures. Thus $\beta$ should
decrease with the increase  of $E$. According to the
Richardson-Zaki relation, the  coefficients $\alpha_{g,p}$
decrease with the increase of  $\rho_{g,p}$ (i.e. void fraction).
In order to mimic this effect on qualitative level we assumed that
$\alpha_{g,p}$ decreases with the increase of $E$ (due to the
increase of maximum density $\rho \sim 1/\beta$ with $E$). Since
the gas is mostly concentrated near the upper electrode and the
precipitate near the bottom,  gas and precipitate are advected in
{\it opposite directions}, i.e. the transport coefficients
$\alpha_{p,g}$ have opposite signs. Since gas is more mobile than
precipitate, we set $D_g \gg D_p$. In the limit $D_g \to \infty$
and for $\alpha_{p,g} =0$ one recovers the description of Ref.
\cite{ar2}.

We assume that vertical vorticity of the liquid
$\Omega_z=\partial_x v_y-\partial_y v_x$ is small in comparison
with the in-plane vorticity. This assumption is justified by the
experimental observation that toroidal vortices create no or very
small horizontal rotation. From  $\Omega_z=0$ one obtains
\begin{equation}
{\bf v} _\perp=-\nabla \phi
\label{phi}
\end{equation}
where $\phi$ is a ``quasipotential''. Substituting Eq. (\ref{phi})
into the continuity equation  $\nabla {\bf v}=0$ one expresses the
quasipotential through vertical velocity
\begin{equation}
\nabla^2 \phi= \partial_z v_z
\label{v_z}
\end{equation}
%The absence of vertical vorticity has one very important
%consequence.

The vertical velocity $v_z$ can be obtained form the corresponding
Navier-Stokes equation
\begin{equation}
\rho_0 (\partial_t v_z + {\bf v} \nabla v_z)=
\nu \nabla^2 v_z -\partial_z p + E_z q
\label{nse}
\end{equation}
where $\rho_0$ is the density of liquid (we set $\rho_0=1$), $\nu$
is the viscosity, $p$ is the pressure, and $q$ is the charge
density. The last term describes electric force acting on charged
liquid. In order to average Eqs. (\ref{v_z}),(\ref{nse}) over the
thickness of the cell $0<z<d$, we assume that $v_z$ is symmetric
with respect to $d/2$ and $v_\perp$ is antisymmetric. Then after
the averaging and taking into account that $\partial_z v_z=0$ at
$z=0,d$ one obtains
\begin{equation}
\partial_t V =
\nu \nabla^2 V -\zeta V -\Delta  p + \langle E_z q\rangle
\label{nse1}
\end{equation}
where $V= d^{-1} \int_0^d v_z dz $, and $\Delta p = p(d) -p(0) $.
Term  $-\zeta V$ accounts for a small dissipation due to friction
between liquid, particles, and the walls of container. Using the
symmetry condition and integrating Eq. (\ref{v_z}) over the lower
half of the cell ($0<z<d/2$) one obtains
\begin{equation}
\nabla^2 \Phi=  a_0 V
\label{v_z1}
\end{equation}
where the constant $a_0  \sim O(1)$ can be scaled away and $\Phi$
is averaged $\phi$. The volume charge density $q \sim - c $ is
negative and proportional to  additive concentration $c$, vertical
component of the electric field $E_z$ depends on  $ E$ and the
local density of particles, $\rho_p + \rho_g$. Since the increase
in the amount of conducting particles  decreases the effective
spacing between the electrodes and, therefore, increases the
apparent electric field, one obtains
\begin{equation}
E_z \approx \frac{E}{1 -(\rho_p+\rho_g)/s} \sim  E + E
(\rho_p+\rho_g)/s \label{Rho}
\end{equation}
where constant $s \approx d/r_0$, $r_0$ is particle diameter.
After applying the divergence operator to the Navier-Stokes
equation one finds that the pressure itself is a functional of
$\langle E q\rangle$. In the most general form it can be written
as $\Delta p = \int K(r -r^\prime ) \langle E_z q\rangle
dr^\prime$. The kernel has  property $\int K dr^\prime =1 $ since
uniform force distribution does not create net flow of the liquid.
The precise form of the kernel is not available on our level of
description. We used the following kernel expressed by its Fourier
transform
\begin{equation}
\hat K (k) = \int \exp[ i {\bf k r} ] K(r) dr = \exp [ -\kappa(E)  k^2 - k^4 d_0^4 ]
\label{kernel}
\end{equation}
This  form of the kernel is justified by the  spatial isotropy
(dependence of $k^2$ only), normalization ($\hat K(0)=\int
K(r^\prime ) dr^\prime =1$) and locality conditions (rapid decays
for $k>1/d_0$, where $d_0$ is the characteristic length of the
order of  electrodes separation). The field dependent factor
$\kappa(E)$ describes experimentally observed transitions from
honeycombs ($\kappa<0$) at small fields to coalescence and
attraction of toroidal vortices ($\kappa>0$) for larger fields.
After combining Eqs. (\ref{nse1}),(\ref{Rho}) and (\ref{kernel})
one obtains

\begin{equation}
\partial_t V =
\nu \nabla^2 V -\zeta V -c E  \int  K_1(r -r^\prime )
(\rho_p+\rho_g) dr^\prime \label{nse2}
\end{equation}
where the Fourier transform  $\hat K_1=1-\exp [ -\kappa(E)  k^2 - k^4 d_0^4 ]$. We also
scaled away $s$.

The  function $f$ has a different structure for low concentrations
 ($c \ll 1 $) and high concentrations. For $c \to 0$ this
function coincides with that derived in Ref. \cite{ar2}:
\begin{equation}
f_1=(\rho_p-\rho_*)\times\left\{
\begin{array}{lll}
\rho_p,& \mbox{if~}& 0\le \rho_p\le \rho_*\\
C_* \rho_g(1-\rho_p),& \mbox{if~}& \rho_*\le \rho_p\le 1  \, .
\end{array}
\right. \label{f1}
\end{equation}
where constants $C_* \sim O(1)$ and $\rho_*(E)$ are discussed in
Ref. \cite{ar2}. For higher values of $c$ the experiment suggests
that both $\rho_g$ and $\rho_p$ tend to some equilibrium value
which is determined by the field $E$. In the linear approximation
it can be written as
\begin{equation}
f_2=c C_2 ( \rho_g - \mu(E,q) \rho_p)
 \label{f2}
\end{equation}
Here $C_2$ characterizes relaxation time towards the equilibrium.
In steady-state $f_2$ enforces the relation between the densities
$\rho_g /\rho_p \to \mu $. In turn, $\mu$ depends on the  field
and local charge distribution $q$.  We set for simplicity
$q=const$. However, convective flows will affect the ions density,
which will shift the gas/precipitate equilibrium, see \cite{sap}.
This effect can be  modelled by including the dependence of $\mu$
on $V$. We used the following expression $\mu= \mu_0 (E) (\tanh(
\mu_1 \mbox{sign}(E) V)+1)$. The particular form of this function
appears to be not relevant. It incorporates the observation that
gas concentration is suppressed by rising flows for $E<0$ due to
excess of negative ions and vise versa. Finally, we used hybrid
form for $f$ valid for arbitrary $c$
\begin{equation}
f = f_2 + \exp[-c/c_0] f_1 \label{c0}
\end{equation}
where $c_0$ is some "crossover" concentration.

{\it Stability of homogeneous precipitate}. In the limit of small
electric fields $E$ we can set $\mu \to 0$, i.e. equilibrium gas
density $\rho_g \to 0$. Then the stability of homogeneous
precipitate (i.e. homogeneous ``Wigner'' crystal state in
experiment) $\rho_p =\bar \rho= const$ can be readily performed
because   the equation for $\rho_g$ splits off and the  analysis
is reduced to Eq. (\ref{nse2}) and Eq. (\ref{con_laws}) for
$\rho_p$. We will focus on the case of high concentration $c \gg
c_0$ and neglect the last term in Eq. (\ref{c0}) (case of $c=0$
was  considered in Ref. \cite{ar2}). In the linear order for
periodic perturbations $V, \rho_p \sim \exp[\lambda t+i k x] $
Eqs. (\ref{con_laws}), (\ref{nse2}) yield
\begin{eqnarray}
\lambda V &=& -(\nu k^2+\zeta) V+ E (1-\exp [ -\kappa(E)  k^2 - k^4 d_0^4 ]) \rho_p  \nonumber \\
\lambda \rho_p &=& -D_p k^2 \rho_p - \alpha_p \beta \bar \rho
(1-\beta \bar \rho) V \label{lin}
\end{eqnarray}
The growth rate $\lambda(k)$  depends  on the sign of the field
$E$, see Figure \ref{figure1}. For negative $E$ and $\kappa<0$ in
a certain parameter range  $\lambda$ is positive  in a narrow band
near the optimal wavenumber $k_0$, which is the hallmark of
so-called short-wave instability. In this regime our solution
indicates formation of stable honeycomb lattice with the scale
determined by $k_0$. For the same parameters, but for $E>0$ the
homogeneous state $\rho_p=\bar \rho$ is stable (dashed line on
Figure \ref{figure1}). This observation is consistent with
experimental fact that field reversal transforms honeycomb to
homogeneous precipitate and vise versa. With the increase of $E$
and $\kappa$ the left edge of the unstable band of $\lambda(k)$
crosses zero and long-wave instability ensues. This regime
corresponds to the coalescence  of clusters.

\begin{figure}[ptb]
\includegraphics[angle=0,width=3.2in]{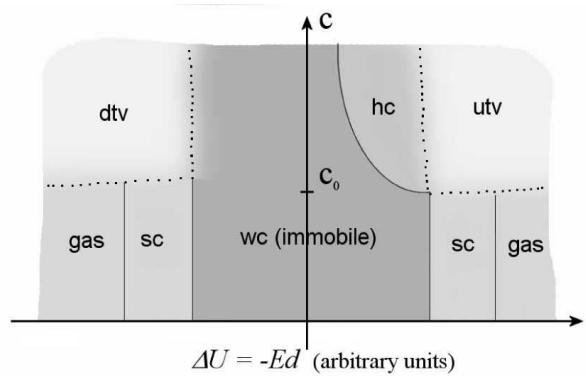}
\caption{Qualitative phase diagram, $\Delta U=- E d$ is applied
voltage (plus on top plate), $c$ is concentration of additive.
Domain {\bf sc} denotes static clusters, {\bf wc} denotes stable
homogeneous precipitate (Wigner crystals), {\bf hc} -honeycombs,
{\bf utv} and {\bf dtv} up/down toroidal vortices.
}\label{figure5}
\end{figure}
\begin{figure}[ptb]
\includegraphics[angle=-90,width=3.3in]{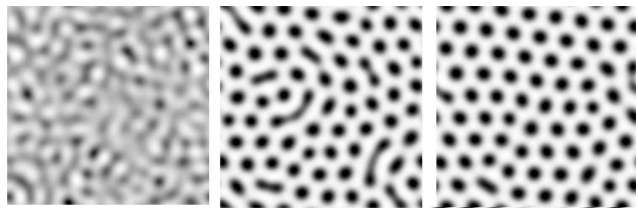}
\caption{ Formation of honeycomb, shown snapshots of $\rho_p$, for
$\bar \rho=0.3, \nu=2, E=-50, \alpha=0.02, \mu_0=0, C_1=5,
 c=1, c_0=0.1, C_*=10, D_p=1, d_0=1, \alpha_p=-\alpha_g=-0.6,
\kappa=-0.1, \beta=2 $, domain of integration $80\times 80$
dimensionless units. Images from left to right: t=10; 400; 2000.
Black corresponds to $\rho_p=0$, white to $\max (\rho_p)$.
%See
%also Movie 1 in \cite{aux}.
}
\label{figure2}
\end{figure}

{\it Qualitative phase diagram} is shown in Figure \ref{figure5}.
The positions of  transition lines are approximate because   field
dependence of $\kappa$, $\mu_0$, $\beta$ is not available at the
moment. To describe the observed phenomenology, we only assumed
that $\kappa$, $\mu_0$, $\rho_*$ monotonically increase with the
increase of $E$, and the functions $\alpha_{g,p},\beta$ decreases
($1/\beta$ is limiting density of the t-vortex which increases
with the increase of $E$). The line $\lambda=0$ for
short-wavelength instability for $E<0$ depicts the transition to
honeycombs. The transition from honeycombs to pulsating rings is
identified as a transition from shortwave to long-wave
instability, which roughly coincides with $\kappa(E)=0$. This
transition is associated with an overall increase of gas
concentration (increase of $\mu(E)$) and decrease of $\beta$
(increase of maximum density of $\rho_p$). The transition from
static clusters to dynamic structures and honeycombs occurs
approximately with the increase of concentration $c$ at $c=c_0$.
For $E>0$  the transition from stable precipitate to ``down
t-vortices'' is approximately given by $\kappa(E)=0$. Note that in
this case there is no distinction between ``immobile'' and
``Wigner crystal'' state because our model does not take into
account dry friction and adhesion of particles to the bottom.

%For small value of $c \ll
%c_0$ the results of Ref. \cite{ar2} were reproduced: formation and
%coarsening of static clusters.

{\it Numerical solution} of Eqs. (\ref{nse2}),(\ref{con_laws}) was
performed by quasi-spectral method based on Fast Fourier
Transformation (FFT). Typically 256$\times$256 FFT harmonics in
periodic boundary conditions were used. For $c>c_0$, $E<0$ and
$\kappa<0$ corresponding to the case of short-wavelength
instability we observed spontaneous formation of honeycomb lattice
from random initial conditions, see Figure \ref{figure2}. For the
same parameters but $E>0$ the homogeneous state $\rho_p=const$
(Wigner crystal)  was stable. With the increase of $E$ and for
$\kappa>0$ (i.e. in the case of long-wavelength instability) we
observed the transition from honeycombs/Wigner crystals to the
regime of cluster attraction and coalescence. In this case EHD
flows accelerate coarsening process of t-vortices dramatically.
The inter-vortex distance $R$ in this regime vs time is shown in
Figure \ref{figure1}, inset. The fitting gives approximately $R
\sim (t_0-t)^{1/2} $. It is qualitatively consistent with
experimental data. This behavior can be anticipated from solution
of Eq. (\ref{v_z1}).  In 2D t-vortices generate localized
distributions of vertical velocity $V$, and, therefore,
logarithmic quasipotentials $\Phi \sim \log|{\bf r} -{\bf r}_i|$,
$r_i$ is the vortex position. Thus, the asymptotic horizontal
velocity will decay as $V_\perp \sim 1/|{\bf r} -{\bf r}_i|$
leading to $R \sim (t_0-t)^{1/2} $.
%The Ref. \cite{sap} reports
%faster coalescence $R\sim (t_0-t)^{1/3}$, although relatively low
%experimental precision makes quantitative comparison difficult.
\begin{figure}[ptb]
\includegraphics[angle=-90,width=3.3in]{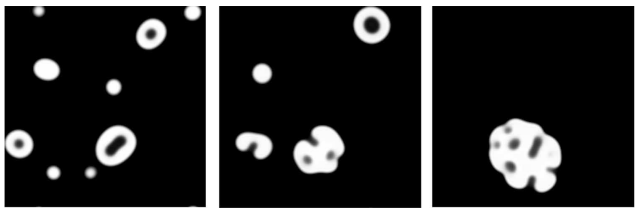}
\caption{Evolution of pulsating rings,  shown snapshots of
$\rho_p$, for $\bar \rho=0.18, \nu=2, E=-70, \alpha=0.02,
\mu_0=0.25, \mu_1=0.5,  C_1=5,
 c=1, c_0=0.1, C_*=10, D_p=1, D_g=5, d_0=1, \alpha_p=-\alpha_g=-0.15,
\kappa=0.1, \beta=0.5 $, domain of integration $80\times 80$
units. Images from left to right: t=450; 1120; 3000.
%See also
%Movies 2 \& 6 in \cite{aux}.
}
\label{figure3}
\end{figure}
\begin{figure}[ptb]
\includegraphics[angle=-90,width=3.3in]{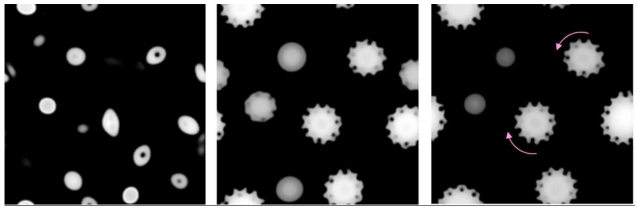}
\caption{Rotating down t-vortices,  shown snapshots of $\rho_p$,
for $\bar \rho=0.18, \nu=2, E=160, \alpha=0.02,
\mu_0=19.5,\mu_1=0.5, C_1=0.45,
 c=1, c_0=0.1, C_*=10, D_p=1, D_g=5, d_0=1,\alpha_p=-\alpha_g=-0.15,
\kappa=0.1, \beta=0.5 $, domain of integration $150\times 150$
units. Images from left to right: t=16; 320; 540.
%See also Movies
%3 \& 7 in \cite{aux}.
}
\label{figure4}
\end{figure}

In the course of coalescence the vortices grow and  become
unstable, producing pulsating rings for $E<0$ (Figure
\ref{figure3}) and rotating objects for $E>0$ (Figure
\ref{figure4}). We believe that the instability  in both cases is
caused by the same mechanism: dynamic coupling between particles
density and vertical flows described by the function $f$. However,
for $E<0$, i.e for utv, the vortices are primarily built of a
low-mobile precipitate phase with a small amount of gas above
them. The instability occurs in the bulk of the vortex, resulting
in formation and breaking up of "gas bubbles" inside the
precipitate. For $E>0$ the vortices are formed by a mobile gas
phase with a small content of precipitate. The instability occurs
at the vortex edge yielding counter-propagating clock/anticlock
waves of shape deformation. Eventually only one wave survives,
creating the effect of rotation.

In our experiments we observed also multi-petal vortices
exhibiting almost solid-state rotation. These multi-petal vortices
obviously generate horizontal flow of the liquid which is
neglected in the framework of our theory. We believe that vertical
vorticity is in fact the consequence rather than the reason of
rotation. It is likely that the vortex shape instability present
in our model eventually triggers rotation of  surrounding fluid
and generates vertical vorticity. Explicit incorporation of the
vertical vorticity  will lead to substantial technical
difficulties and likely will not change the qualitative features.

In conclusion, we develop phenomenological continuum theory of
pattern formation and self-assembly of metallic microparticles
immersed in a poorly conducting liquid. This theory  reproduces
primary patterns observed in the experiment, and leads to an
interesting prediction of the relation of  rotation with the
vortex edge instability. The parameters of the model can be
extracted from molecular dynamics simulations and generalization
of  ``leaky dielectric model'' \cite{leaky} and validated by
experiments. We are grateful to B. Meerson, Y. Tolmachev and W.-K.
Kwok for useful discussions. This research was supported by the US
DOE, grant W-31-109-ENG-38.


\begin{thebibliography}{99}


\bibitem{hayward} R.C.Hayward, D.A. Saville,  and I.A. Aksay, Nature {\bf 404},
56(2000); M. Trau et al, Nature {\bf 374}, 437 (1995);  S.-R. Yeh,
M. Seul, and B.I. Shraiman, Nature {\bf 386}, 57 (1997).
\bibitem{chang}R. B. M. Schasfoort et al, Science, {\bf 286}, 942 (1999); A.R. Minerick, A.E. Ostafin, and
H.-C. Chang, Electrophoresis, {\bf 23}, 2165 (2002)
\bibitem{ar1}  I.S. Aranson et al,   Phys. Rev. Lett. {\bf 84}, 3306 (2000)
\bibitem{ar2} I.S. Aranson, B. Meerson, P.V. Sasorov,  and V.M. Vinokur,
Phys. Rev. Lett. {\bf 88}, 204301 (2002)
\bibitem{sap}
M. V. Sapozhnikov, Y. V. Tolmachev, I. S. Aranson, and W.-K. Kwok,
\prl {\bf  90}, 114301 (2003)
\bibitem{richardson} J.F.  Richardson and W.N. Zaki, Trans. Inst.
Chem. Eng. {\bf 32}, 35 (1954); C.Y. Wen and Y.H. Yu, Chem. Eng.
Prog. Symp. Ser. {\bf 62}, 100 (1966)
%\bibitem{aux} See EPAPS Document No.  for movies illustrating dynamics of
%patterns. A direct link to this document may be found in the
%online article's HTML reference section. The document may also be
%reached via the EPAPS homepage (http://www.
%aip.org/pubservs/epaps.html) or from ftp.aip.org in the directory
%/epaps/. See the EPAPS homepage for more information.
\bibitem{leaky} D.A. Saville, Annu. Rev. Fluid Mech, {\bf 29}, 27
(1997)



\end{thebibliography}
\end{document}